\documentclass[twoside,reqno]{HERON}
\usepackage{epsfig,cite,colordvi}
\usepackage{graphicx}
\usepackage{url}
\usepackage{amssymb,amsmath,amscd,epsf}
\usepackage{times}
\usepackage{makeidx}
\def\Vec#1{\mbox{\boldmath $#1$}}

\begin{document}

\title{Universality of Short-Range Correlations
  in One- and Two-Nucleon Momentum Distributions of Nuclei}

\runningheads{Universality of Short-Range Correlations
  in One- and Two-Nucleon Momentum Distributions of Nuclei}{M. Alvioli}

\begin{start}

\author{M. Alvioli}{1,2}

\index{Alvioli, M.}

\address{ECT$^\star$, European Centre for Theoretical Studies in Nuclear Physics 
  and Related Areas, Strada delle Tabarelle 286, I-38123 Villazzano (TN) Italy}{1}
\address{CNR-IRPI, Istituto di Ricerca per la Protezione
  Idrogeologica - Via Madonna Alta, 126 - I-06128 Perugia (PG) Italy}{2}

\begin{Abstract}
Universality of short range 
correlations has been investigated both in coordinate and in momentum space, by means of
one-and two-body densities and momentum distributions. In this contribution we discuss
one- and two-body momentum distributions across a wide range of nuclei and their
common features which can be ascribed to the presence of short range correlations.
Calculations for few-body nuclei, namely $^3$He and $^4$He, have been performed using exact 
wave functions obtained with Argonne nucleon-nucleon interactions, while the linked
cluster expansion technique is used for medium-heavy nuclei. 
The center of mass motion of a nucleon-nucleon pair in the nucleus, embedded in
the full two-body momentum distribution $n^{NN}(\Vec{k}_{rel}, \Vec{K}_{CM})$, 
is shown to exhibit the universal behavior predicted by the two-nucleon correlation model, 
in which the nucleon-nucleon pair moves inside the nucleus as a deuteron in a mean-field. 
Moreover, the deuteron-like spin-isospin (ST)=(10) contribution to the pn two-body momentum 
distribution is obtained, and shown to exactly scale to the deuteron momentum distribution. 
Universality of correlations in two-body distributions is cast onto the one-body distribution 
$n(k_1)$, obtained by integration of the two-body $n^{NN}(\Vec{k}_1, \Vec{k}_2)$: in particular, 
the high momentum part of $n(k_1)$ exhibits the same pattern for all considered nuclei, in favor 
of a universal character of the short range structure of the nuclear wave function.
Perspectives of this work, namely the calculation of reactions involving light and complex nuclei 
with realistic wave functions and effects of Final State Interactions (FSI), investigated by 
means of distorted momentum distributions within the Glauber multiple scattering approach, are 
eventually discussed.
\end{Abstract}
\end{start}

We report on recent developments on the investigation of Short Range Correlations (SRC) in nuclei,
whose common properties have recently been studied by a few authors in light 
\cite{Sargsian:2005ru,Schiavilla:2006xx,Alvioli:2011aa,Feldmeier:2011qy,Vanhalst:2011es} and medium-heavy 
\cite{Piasetzky:2006ai,Alvioli:2007zz,Alvioli:2007vv,Alvioli:2012qa} nuclei using realistic nuclear wave
functions. Common properties among different nuclei have been investigated both in coordinate and
momentum space using various methods and Nucleon-Nucleon (NN) interaction potentials for generating 
nuclear wave functions. Such an effort from the theoretical side was motivated by the recent observation 
of NN SRC in different experiments (reviewed in \cite{Arrington:2011xs}): two nucleon at high 
four-momentum transfer with 
protons and electrons, namely A(p,ppN)X of Ref. \cite{Tang:2002ww} and A(e,e$^\prime$pN)X of 
Refs. \cite{Shneor:2007tu,Subedi:2008zz}, and inclusive electron scattering A(e,e$^\prime$)X 
of Refs. \cite{Egiyan:2005hs,Fomin:2011ng}, have provided evidence that NN SRC: 
i) exist in the ground-state
wave function of nuclei; ii) can be detected in different reactions, using different projectiles
and final states, suggesting some universal character, though direct measurement of two nucleons 
in a correlated pair have been performed only in the carbon nucleus; iii) are dominated by tensor 
correlations, which is suggested by the small observed fraction of correlated proton-proton pair 
as compared to the proton-neutron ones, and originating from the two-body tensor operator acting 
between pairs of nucleons in a state with spin S=1 and isospin T=0. Universality patterns
observed in momentum space in the high-momentum region, correspond to similar
behavior in coordinate space
\cite{Baldo:1900zz,CiofidegliAtti:2010xv,Roth:2010bm,Feldmeier:2011qy,Bogner:2012zm}. 
Moreover, the existence of SRC in nuclei has been related to nuclear EMC effect and poses the 
question of the validity of many  of the proposed models of the EMC effect
\cite{Weinstein:2010rt,Frankfurt:2012qs}.

The dominance of the tensor part of the NN interaction and its effects on the two-body momentum
distributions were clearly described from different theoretical groups for light 
\cite{Schiavilla:2006xx} and medium-heavy nuclei \cite{Alvioli:2007zz,Alvioli:2007vv}. 
Moreover, the pn vs. pp content of the two-body momentum distributions of a nucleon pair with
zero center of mass was calculated in Ref. \cite{Alvioli:2007zz}, and found to be compatible 
with the observed ratio of pn over pp correlated pairs. Nevertheless, a consistent and a 
quantitative description of the detected pairs in the experimental kinematics should use the full 
information contained in the wave functions, so that the center of mass momentum of the pair is 
allowed to be different from zero, and should take into account the complex FSI suffered by the 
two nucleons before leaving the nucleus and being detected.

The calculation of the center of mass dependence of two-body momentum distributions for light
\cite{Alvioli:2011aa} and medium-heavy nuclei \cite{seattle09,miami10} has been performed with 
exact wave functions \cite{Kievsky:1992um,Akaishi}, in the first case, while it relies on the
linked cluster expansion method presented in Refs. \cite{CiofidegliAtti:1999zz,Alvioli:2005cz}, 
in the case of A$\ge$12.
The linked cluster expansion formalism makes use of variationally determined correlation
functions coupled to the spin-isospin dependent operators adopted by realistic interactions.
The correlation function approach is common to several methods, and it is used to describe both 
finite nuclei 
\cite{Pieper:1992gr,Geurts:1996zz,Suzuki:2008cy,Horiuchi:2007ww,AriasdeSaavedra:2007qg} 
and nuclear matter 
\cite{Pandharipande:1979bv,Benhar:1989aw,Baldo:2012zz,Rios:2006mn,Carbone:2011wk,Baldo:2012nh}.
The linked cluster expansion method allows the computational demand to be reduced with respect to 
exact methods, so that the full one- and two-body,
diagonal and non-diagonal density distributions can be calculated, along with the individual 
contributions of given spin and isospin states.

The distributions discussed in this contribution, for all the nuclei we considered, are obtained by 
computing the expectation value of suitable operators on the ground-state wave function. The non-diagonal 
one-body density is obtained as:
\begin{equation}
\label{obddef}
\rho^{(1)}_N(\Vec{r}_1;\Vec{r}^\prime_1)\,=\,\sum_{\{\sigma,\tau\}}\int\prod^A_{j=2}d\Vec{r}_jd\Vec{r}^\prime_j
\Psi^\star(\{\Vec{x}\})\,\hat{\rho}^{(1)}\,\Psi(\{\Vec{x}^\prime\})\,,
\end{equation}
while the two-body density is:
\begin{equation}
\label{tbddef}
\rho^{(2)}_{NN}(\Vec{r}_1,\Vec{r}_2,;\Vec{r}^\prime_1,\Vec{r}^\prime_2)\,=\,\sum_{\{\sigma,\tau\}}\int\prod^A_{j=3}
d\Vec{r}_jd\Vec{r}^\prime_j
\Psi^\star(\{\Vec{x}\})\,\hat{\rho}^{(2)}\,\Psi(\{\Vec{x}^\prime\})\,,
\end{equation}
where $\{\sigma\}$ and $\{\tau\}$ stand for the spin and isospin degrees of freedom, respectively,
of the A nucleons; $\Vec{x}=\{\Vec{r},\sigma,\tau\}$, and the primed variables stands for fluctuation
of the spatial part $\Vec{r}$, while the spin and isospin variables of the corresponding particle
are fixed. We use a Slater determinant $\Phi$ of shell model single particle wave function for $\Psi$,
and the correlations are implemented, by means of an operator $\hat{F}$ containing the correlation 
functions and corresponding state-dependent operators \cite{Alvioli:2005cz}:
$\Psi(\Vec{x}_1,\cdots,\Vec{x}_A)\,=\,\hat{F}(\Vec{x}_1,\cdots,\Vec{x}_A)\,\Phi(\Vec{x}_1,\cdots,\Vec{x}_A)$,
in such a way that the final wave function is totally antisymmetric.

The operators appearing in Eqs. (\ref{obddef}) and (\ref{tbddef}) are respectively defined as:
\begin{eqnarray}
\label{obdop}
\hat{\rho}^{(1)}\hspace{-0.3cm}&=&\hspace{-0.3cm}\sum^A_i\delta(\Vec{r}_i-\tilde{\Vec{r}}_1)\,\delta(\Vec{r}^\prime_i-\tilde{\Vec{r}}^\prime_1)
\,\prod^A_{k\neq i}\,\delta(\Vec{r}_k-\Vec{r}^\prime_k)\,;\\
\label{tbdop}
\hat{\rho}^{(2)}\hspace{-0.3cm}&=&\hspace{-0.3cm}\sum^A_{i<j}\delta(\Vec{r}_i-\tilde{\Vec{r}}_1)\,\delta(\Vec{r}_j-\tilde{\Vec{r}}_2)
\,\delta(\Vec{r}^\prime_i-\tilde{\Vec{r}}^\prime_1)\,\delta(\Vec{r}^\prime_j-\tilde{\Vec{r}}^\prime_2)
\hspace{-0.1cm}\prod^A_{k\neq i,j}\hspace{-0.1cm}\delta(\Vec{r}_k-\Vec{r}^\prime_k)\,.
\end{eqnarray}

\begin{figure}[!htp]
  \centerline{
    \includegraphics[scale=0.77]{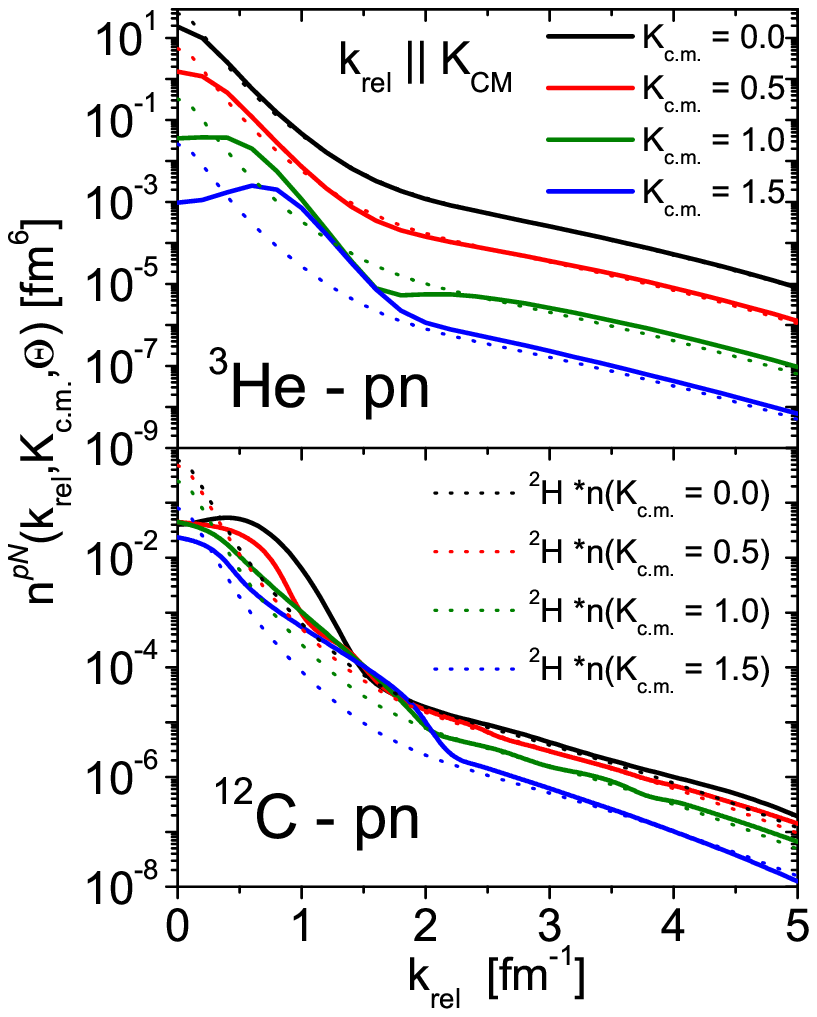}
    \hspace{-1.66cm}
    \includegraphics[scale=0.77]{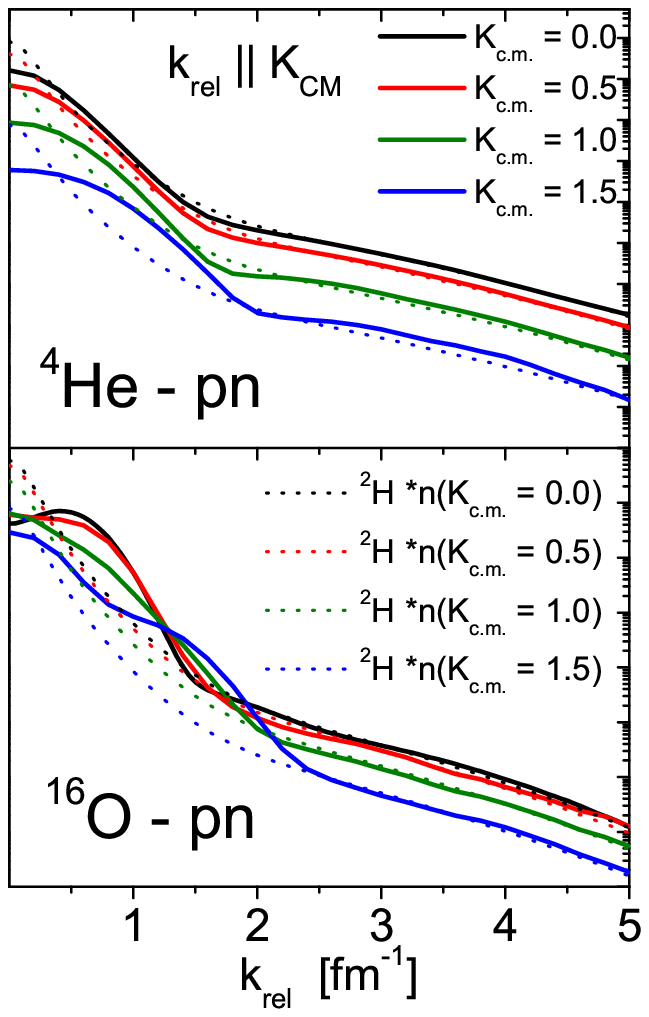}}
    \caption[]{The relative two-body momentum distribution for a proton-neutron
      pair in $^3$He and $^4$He (top panels; after Ref. \cite{Alvioli:2011aa}) 
      and in $^{12}$C and $^{16}$O (lower panels; after Refs. \cite{seattle09,miami10})
      at fixed values $K_{CM}$=0.0, 0.5, 1.0 and 1.5 fm$^{-1}$ of the center of mass
      of the pair. The various curves are compared to the deuteron distribution,
      multiplied by the value of $n(K_{CM})=\int d\Vec{k}_{rel}n(k_{rel},\Vec{K}_{cm})$
      at the corresponding values of $K_{CM}$; see Ref. \cite{Alvioli:2011aa} for details
      and additional calculations.}
    \label{fig1}
\end{figure}

The one- and two-body momentum distributions are obtained from Eqs. (\ref{obddef}) and (\ref{tbddef})
by proper Fourier transformations:
\begin{eqnarray}
\label{obmd}
n^N(k_1)\hspace{-0.0cm}&=&\hspace{-0.0cm}\frac{1}{(2\pi)^3}\int d\Vec{r}_1d\Vec{r}^\prime_1
\,e^{i\,\Vec{k}_1\cdot(\Vec{r}_1-\Vec{r}^\prime_1)}\,\rho^{(1)}_N(\Vec{r}_1;\Vec{r}^\prime_1)\,,\\
n^{NN}(\Vec{k}_1,\Vec{k}_2)\hspace{-0.0cm}&=&\hspace{-0.0cm}\frac{1}{(2\pi)^6}\int d\Vec{r}_1d\Vec{r}_2
d\Vec{r}^\prime_1d\Vec{r}^\prime_2\,e^{i\,\Vec{k}_1\cdot(\Vec{r}_1-\Vec{r}^\prime_1)}\,\cdot\nonumber\\
\label{tbmd}&&\hspace{2cm}\cdot\,e^{i\,\Vec{k}_2\cdot(\Vec{r}_2-\Vec{r}^\prime_2)}
\,\rho^{(2)}_{NN}(\Vec{r}_1,\Vec{r}_2;\Vec{r}^\prime_1,\Vec{r}^\prime_2)\,;
\end{eqnarray}
the two-body momentum distributions can be conveniently rewritten in terms of the relative and
center of mass momenta as follows:
\begin{eqnarray}
n^{NN}(\Vec{k}_{rel},\Vec{K}_{CM})\hspace{-0.0cm}&=&\hspace{-0.0cm}\frac{1}{(2\pi)^6}\int 
d\Vec{r}d\Vec{R}d\Vec{r}^\prime d\Vec{R}^\prime
\,e^{i\,\Vec{k}_{rel}\cdot(\Vec{r}-\Vec{r}^\prime)}\,\cdot\nonumber\\
\label{tbmd2}&&\hspace{1cm}\cdot\,e^{i\,\Vec{K}_{CM}\cdot(\Vec{R}-\Vec{R}^\prime)}
\,\rho(\Vec{r},\Vec{R};\Vec{r}^\prime,\Vec{R}^\prime)\,,
\end{eqnarray}
were $\Vec{r}=\Vec{r}_1-\Vec{r}_2$, $\Vec{R}=(\Vec{r}_1+\Vec{r}_2)/2$ (and analogous definitions
for the primed vectors) and $\Vec{k}_{rel}=(\Vec{k}_1-\Vec{k}_2)/2$, $\Vec{K}_{CM}=\Vec{k}_1+\Vec{k}_2$.
The partial spin and isospin contributions to the quantities above have been obtained by complementing 
the density operators of Eq. (\ref{tbdop}) by the projection operators on specific total spin S and 
isospin T of the ``1,2'' NN pair; extended definitions, normalization conditions and relations between 
the various formulas of Eqs. (\ref{obddef}) and (\ref{tbddef}) are given in Ref. \cite{Alvioli:2012qa}.
It is worth mentioning that actual calculations of two-body momentum distributions, which are obtained 
by sampling the one- and two-body densities after multidimensional integration of the few-body wave 
functions, in the case of $^3$He and $^4$He, and of the corresponding linked cluster expansion formulas 
given in details in Refs. \cite{Alvioli:2005cz,massitesi}, for complex nuclei, are the following. In the 
case of the relative two-body distribution, obtained by integrating Eq. (\ref{tbmd2}) over the center of 
mass momentum, we use the following coordinates: $\Vec{x}=\Vec{r}-\Vec{r}^\prime$, 
$\Vec{t}=\frac{1}{2}\,\left(\Vec{r}+\Vec{r}^\prime\right)$, $\Vec{s}=\Vec{R}-\Vec{R}^\prime$
and $\Vec{w}=\frac{1}{2}\,\left(\Vec{R}+\Vec{R}^\prime\right)$.
Performing the coordinate transformation into Eq. (\ref{tbmd2}) and integrating over $K_{CM}$,
we have:
\begin{eqnarray}
n^{NN}_{rel}(k_{rel})&=&\int d\Vec{K}_{CM}\,n^{NN}(\Vec{k}_{rel},\Vec{K}_{CM})\,=\nonumber\\
&=&\frac{1}{(2\pi)^3}\int
d\Vec{x}\,d\Vec{t}\,d\Vec{s}\,d\Vec{w}\,
e^{i\,\Vec{k}_{rel}\cdot\Vec{x}}\,
\rho^{(2)}_{NN}(\Vec{x},\Vec{t},\Vec{s},\Vec{w})\,=\nonumber\\
&=&\frac{1}{(2\pi)^3}\int
d\Vec{x}\,e^{i\,\Vec{k}_{rel}\cdot\Vec{x}}\,\rho^{(2)}_{NN}(x)\,=\nonumber\\
&=&\label{nrelkrel}\frac{1}{2\pi^2}\int^\infty_0
dx\,x\,\frac{\sin{k_{rel}\,x}}{k_{rel}}\,\rho^{(2)}_{NN}(x)\,,
\end{eqnarray}
where we choose $\Vec{k}_{rel}$ along a given direction that we define as the $z$ axis and with, 
obviously:
\begin{equation}
\rho^{(2)}_{NN}(x)\,=\,\int d\Vec{t}\,d\Vec{s}\,d\Vec{w}\,
\rho^{(2)}_{NN}(\Vec{x},\Vec{t},\Vec{s},\Vec{w})\,,
\end{equation}
with $x=|\Vec{x}|$. Analogously, we have
\begin{equation}\label{ncmkcm}
n^{NN}_{CM}(K_{CM})\,=\,
\frac{1}{2\pi^2}\int^\infty_0 ds\,s\,\frac{\sin{K_{CM}\,s}}{K_{CM}}\,\rho^{(2)}_{NN}(s)\,.
\end{equation}
In the case we want to fix a few (\textit{small}) values of $\Vec{K}_{CM}$, and plot 
the resulting $k_{rel}$ distribution, we have to choose the relative orientation of 
$\Vec{k}_{rel}$ and $\Vec{K}_{CM}$. If we choose both of the them along the same direction, 
for example $z$, Eq. (\ref{tbmd2}) becomes
\begin{eqnarray}
n^{NN}_\parallel(\Vec{k}_{rel},K_{CM})&=&\nonumber\\
&&\hspace{-2cm}=\,\frac{1}{(2\pi)^6}\,Re\int
d\Vec{x}\,d\Vec{t}\,d\Vec{s}\,d\Vec{w}\,
e^{i\,\Vec{K}_{CM}\cdot\Vec{s}}\,
e^{i\,\Vec{k}_{rel}\cdot\Vec{x}}\,
\rho^{(2)}_{NN}(\Vec{x},\Vec{t},\Vec{s},\Vec{w})\,=\nonumber\\
&&\hspace{-2cm}=\,\frac{1}{(2\pi)^6}\,Re\int
d\Vec{x}\,d\Vec{t}\,d\Vec{s}\,d\Vec{w}\,
e^{i\,K_{CM}\,s_z}\,e^{i\,k_{rel}\,x_z}\,
\rho^{(2)}_{NN}(\Vec{x},\Vec{t},\Vec{s},\Vec{w})\,=\nonumber\\
&&\label{realSZ}\hspace{-2cm}=\,\frac{1}{(2\pi)^6}\int^\infty_{-\infty}\,dx_z\,ds_z
\,\cos{\left(k_{rel}\,x_z + K_{CM}\,s_z\right)}\,\rho^{(2)}_{NN}(x_z,s_z)\,,
\end{eqnarray}
where, in this case, we have to sample a two-dimensional
$\rho^{(2)}_{NN}(x_z,s_z)$, defined as
\begin{equation}
\rho^{(2)}_{NN}(x_z,s_z)\,=\,\int
d\Vec{t}d\Vec{w}\,dx_x\,dx_y\,ds_x\,ds_y\,\rho^{(2)}_{NN}(\Vec{x},\Vec{t},\Vec{s},\Vec{w})\,,
\end{equation}
and we have taken the real part in Eq. (\ref{realSZ}). Similarly, we can calculate the
distribution in the case of the relative momentum perpendicular to the center of mass momentum:
\begin{eqnarray}
\label{realSX}
n^{NN}_\perp(\Vec{k}_{rel},K_{CM})\,=\,\int\,\frac{dx_z\,ds_x}{(2\pi)^6}
\,\cos{\left(k_{rel}\,x_z + K_{CM}\,s_x\right)}\,\rho^{(2)}_{NN}(x_z,s_x)\,,
\end{eqnarray}
with $\rho^{(2)}_{NN}(x_z,s_x)$ defined accordingly; in this case, we choose $k_{rel}$ along
the $z$ axis, and $K_{CM}$ along the $x$ axis. Results for the parallel case, Eq. (\ref{realSZ}),
are shown in Fig. \ref{fig1}, for various nuclei. We also compare these results with the predictions 
of the Two-Nucleon Correlation (TNC) model, where the relative distribution is taken as the deuteron's 
one, and the center of mass Gaussian distribution of the model has been replaced by actual values of 
$n^{pn}_{CM}(K_{CM})$, the two-body momentum distributions integrated over the relative momentum, 
Eq. (\ref{ncmkcm}).

The general conclusions that can be drawn about the two-body momentum distribution 
$n^{NN}(\Vec{k}_{rel},\Vec{K}_{CM})$ are: i) at $K_{CM}=0$, 
\textit{i.e.} for nucleons moving exactly back-to-back, in the so-called correlated region 
1.5 fm$\lesssim k_{rel}\lesssim$ 3.0 fm, the distribution as a function of 
$k_{rel}$, $n^{NN}(k_{rel},K_{CM}=0)$, has a shape which is similar to deuteron's S wave, in the
proton-proton pair case, and similar to the deuteron's D wave, in the proton-neutron pair case
\cite{Schiavilla:2006xx}; ii) the node in the S wave is partially filled increasing A from $^3$He
and $^4$He to $^{12}$C and $^{16}$O \cite{Alvioli:2007vv,Alvioli:2007zz,Alvioli:2011aa,seattle09,miami10};
iii) for non-zero values of $K_{CM}$, we find a decreasing high-momentum tail, with respect to the 
$K_{CM}=0$ case \cite{Alvioli:2007vv,Alvioli:2007zz,Alvioli:2011aa,seattle09,miami10}; 
iii) the high-momentum tail of the proton-neutron distribution can be checked against
TNC model of Ref. \cite{CiofidegliAtti:1991mm}, where the deuteron distribution was convoluted with a 
Gaussian center of mass dependence to obtain a model 
two-body momentum distribution: the comparison shows that up to moderate values of $K_{CM}$, the deuteron 
shape is reproduced by realistic calculations \cite{Alvioli:2007vv,Alvioli:2007zz,Alvioli:2011aa,seattle09,miami10}; 
iv) in the region of small values of $K_{CM}$ and 1.5 fm$\lesssim k_{rel}\lesssim$ 3.0 fm, the proton-neutron 
distributions at $K_{CM}$\textgreater0 can be obtained from the one at $K_{CM}$=0 scaled by a factor which is 
given by $n^{pn}_{CM}(K_{CM})$ of Eq. (\ref{nrelkrel}), and depends only on the modulus of $K_{CM}$ \cite{Alvioli:2011aa} 
(cfr. Fig. \ref{fig1}); v) starting from $^4$He the Gaussian approximation of Ref. \cite{CiofidegliAtti:1991mm} 
seems to be a good one, and the agreement with the Gaussian shape is better in larger nuclei \cite{Alvioli:2011aa}; 
vi) the comparison of $n^{NN}(\Vec{k}_{rel},\Vec{K}_{CM}=0)$ with the deuteron distribution is an approximate 
one: it was found that extracting from the total distribution the only contribution due to proton-neutron 
pairs in a spin 0 and isospin 1 state, which is possible in our formalism both in the light nuclei as well 
as in the medium-heavy nuclei case, the agreement becomes quantitative and the ratio of the nucleus to 
deuteron distribution is a constant in the correlation region \cite{Alvioli:2011aa}. Moreover, we recently 
extended our analysis to the one-body momentum distributions case in great detail \cite{Alvioli:2012qa}, 
showing that similar conclusions can be drawn as in the two-body distributions case, and that the TNC model 
of Ref. \cite{CiofidegliAtti:1991mm} can be effectively be updated exploiting the outcomes of the described
realistic calculations, in view of the possible experimental measurement of correlated pairs in a nucleus 
in different spin and isospin states as well as carrying arbitrary center of mass momentum.

Such a detailed picture of SRC is experimentally limited to the $^{12}$C nucleus, 
where one- and two-nucleon knockout were investigated in the correlation region,
while correlations have been studied in a number of nuclei in inclusive electron
scattering at $x=Q^2/2M_N\nu$\textgreater1 and low $\nu$. 
Experimental information on the center of mass momentum of a correlated pair in $^{12}$C
from Ref. \cite{Tang:2002ww} was compared with predictions made long before within the
TNC model \cite{CiofidegliAtti:1995qe}, finding quantitative agreement; nevertheless,
as realistic calculations such as those presented in this report have become available
for whole range of nuclei, it is now possible to go beyond the model predictions
\cite{Alvioli:2011aa,Alvioli:2012qa}.
Additional experimental information on the dependence of SRC on the center of mass 
of the pair and its isospin could be compared with theoretical calculations as the
ones presented in this report.
Our calculations of two-body distributions and their interpretation in terms of TNC model 
for correlated pairs allows one to single out the regions on the $\Vec{k}_{rel}$, $\Vec{K}_{CM}$ 
plane in which the various contributions to 
$n^{NN}(\Vec{k}_{rel},\Vec{K}_{CM})$ are dominant. The region of small momenta
is clearly dominated by the shell model contribution, while the region of high relative
and small center of mass momenta of the pair is dominated by the two-nucleon
correlations. As the center of mass momentum of the pair increase, configurations 
with three nucleons with vanishing total momentum start to dominate, and this is the 
region in which three-body correlations effects should be investigated, as it is 
argued in Ref. \cite{Alvioli:2011aa}. While experimental information on the relevance 
of three-nucleon correlations are still missing, evaluation of the
contributions due to particular configurations of three nucleons whose total momentum
is small seems to be feasible in the range of light to medium-heavy nuclei.

\begin{figure}[!t]
  \centerline{
    \includegraphics[scale=0.88]{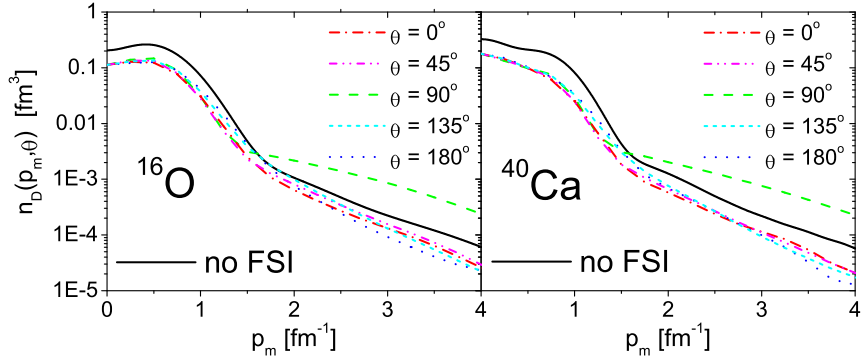}}
    \caption[]{The distorted momentum distribution $^{16}$O and $^{40}$Ca
      calculated within the Glauber model for the FSI of the knocked out
      proton in an A(e,e$^\prime p$)X process and correlated nuclear wave
      functions, as proposed in Ref. \cite{CiofidegliAtti:1999zz}.}
    \label{fig2}
\end{figure}

As we already mentioned, a full calculation of the processes used to directly observe SRC 
in $^{12}$C is still missing. A consistent calculation should take into account the realistic 
initial state
and full FSI between the knocked out nucleons and the residual nucleons, in the same way 
in which calculations for light nuclei has been performed (see
\cite{CiofidegliAtti:2005qt,Sargsian:2005ru,CiofidelgiAtti:2007qu,CiofidegliAtti:2008ux,Alvioli:2009zy,Palli:2009it,AlvarezRodriguez:2010nb}) 
and have proven to reproduce cross section data.
Inclusion of SRC in the initial state can be done by means of the notion of two-nucleon overlap
function, the overlap between the initial A-body state and the (A-2)-body plus the two knocked 
out nucleons final state 
\cite{Simpson:2010yg,Simpson:2012hx,Antonov:1998gf,Kadrev:2002pi,Giusti:1991bs,Cosyn:2009bi}.
In particular,
the approach of Ref. \cite{Kadrev:2002pi} where the authors perform calculations for two-proton
knockout reactions with Jastrow correlations implemented with a linked cluster expansion similar
to the one we used in our work, seems to be particularly suitable for an extension to realistically 
correlated wave functions calculated within our formalism. As a matter of fact, the Jastrow 
correlation functions misses the complexity of the full correlation structure induced by realistic 
potentials: in particular, they lack the tensor interaction which is a key ingredient for distinguishing
basic properties of proton-proton as compared to proton-neutron correlations. For this reason
an extension of the calculations of Ref. \cite{Kadrev:2002pi}, in which only the reaction
$^{16}$O(e,e$^\prime$pp)$^{14}$C(g.s.) C was discussed, with a fully correlated 
wave function, would be particularly meaningful. As far as FSI are concerned, it should
be mentioned that the use of generalized Glauber multiple scattering approach for the rescattering
of the knocked out nucleons in the final state in conjunction with realistic wave functions, in
the case of light nuclei, proved to be very successful when comparing to experimental data, and
basic features concerning the behavior of the correlated pair are reproduced by the TNC model. 
Calculations of distorted momentum distributions within the framework of the Glauber theory, and using 
correlated initial states for medium-heavy nuclei, can be performed in our formalism 
\cite{CiofidegliAtti:1999zz,massitesi,Alvioli:2003hv,Alvioli:2008as} 
and suggest a common pattern across the range of nuclei we have investigated; partial results 
for complex nuclei, obtained with an extension the formalism of Ref. \cite{CiofidegliAtti:1999zz},
are shown in Fig. \ref{fig2} (see \cite{massitesi}). The figure shows the quantity
\begin{equation}
\label{dnk}
n_D(p_m,\theta)\,=\,\frac{1}{(2\pi)^3}\int d\Vec{r}_1d\Vec{r}^\prime_1
\,e^{i\,\Vec{p}_{m}\cdot(\Vec{r}_1-\Vec{r}^\prime_1)}
\,\rho^{(1)}_D(\Vec{r}_1;\Vec{r}^\prime_1)\,.
\end{equation}
\noindent The distorted momentum distribution of Eq. (\ref{dnk}) correspond to a process
A(e,e$^\prime$p)X, with three-momentum transfer $q$ and missing momentum 
$\Vec{p}_{m}$; $\theta$ is the angle between the missing momentum and the
direction of propagation of the knocked out proton. This direction is singled out 
by the Glauber operator $\hat{S}$ contained in the distorted density: 
%$\rho^{(1)}_D(\Vec{r}_1,\Vec{r}^\prime_1)$
\begin{equation}
\rho^{(1)}_D(\Vec{r}_1;\Vec{r}^\prime_1)\,=\,\sum_{\{\sigma,\tau\}}\int\prod^A_{j=2}
d\Vec{r}_jd\Vec{r}^\prime_j\,e^{i\Vec{p}_m\cdot(\Vec{r}_1-\Vec{r}^\prime_1)}\,
\Psi^\star(\{\Vec{x}\})\,\hat{S}^\dagger\,\hat{\rho}^{(1)}\,\hat{S}\,\Psi(\{\Vec{x}^\prime\})\,,
\end{equation}
and causes the original momentum distribution of the nucleon in the nucleus to be 
distorted and anisotropic, even if the target nucleus is a spherical one.
As expected, the effect of FSI is different at the different values of the angle, 
and a similar pattern is exhibited by calculations at same angle in different nuclei,
despite the different methods used to obtain the wave functions.
A detailed comparison between the results for finite nuclei and the deuteron, and the 
extent to which similarities can be ascribed to universality of FSI within a correlated
pair, is under investigation and will be reported in a separate publication.

\section*{Acknowledgments}
We thank CASPUR for the grant SRCnuc3 - \textit{Short Range Correlations in nuclei}, 
within the Standard HPC grants 2012 programme for providing computer time for the
calculations presented in this Proceedings and related publications.

\end{document}